\documentclass[aps,prb,reprint,groupedaddress]{revtex4-2}

\usepackage{amssymb}
\usepackage{amsmath}
\usepackage{xcolor}
\usepackage{graphicx}
\usepackage{epsfig}
\usepackage{subfigure}

\begin{document}

\title{Giant nonlinearity of kinetic inductance in superconducting hybrid strip}

\author{D. Yu. Vodolazov$^{1,2}$}
\email{vodolazov@ipmras.ru}

\author{Yu. P. Korneeva$^1$}

\author{M. A. Dryazgov$^1$}

\author{S. S. Ustavschikov$^{1,2}$}

\author{N. V. Porokhov$^1$}

\author{N.~N.~Osipov$^1$}

\author{I. Yu. Pashenkin$^2$}

\author{A.A. Korneev$^{1,3}$}

\author{M. A. Tarkhov$^1$}

\affiliation{$^1$Institute of Nanotechnology of Microelectronics
of the Russian Academy of Sciences, 115487, Moscow, Nagatinskaya
16A, build. 11, Russia}

\affiliation{$^2$Institute for Physics of Microstructures, Russian
Academy of Sciences, 603950, Nizhny Novgorod, GSP-105, Russia}

\affiliation{$^3$HSE University, 101000, Moscow, Myasnitskaya 20,
Russia}

\date{\today}

\begin{abstract}

We demonstrate giant nonlinearity of the kinetic inductance $L_k$
in a superconducting NbN/Al hybrid strip. In contrast to ordinary
superconductors, where the maximum observed current-induced
variation of $L_k$ does not exceed tens of percent, we observe a
50-fold change in $L_k$. Physically, this effect is connected with
the proximity-induced superconductivity in the Al layer and the
specific choice of parameters for which the dependence of the
superconducting current $I$ on the Cooper pair momentum $\hbar q$
exhibits a plateau at a current $I^*$ {\it well below} the
depairing current. Formally, this leads to $L_k \sim dq/dI \to
\infty$ as $I \to I^*$. We believe that our results open a new
route for creating the next generation of highly sensitive
detectors of electromagnetic radiation, calorimeters, parametric
amplifiers, and magnetometers.

\end{abstract}

\maketitle

The kinetic inductance $L_k$ of a superconductor arises from the
inertial mass of Cooper pairs. Unlike geometric inductance, which
stores energy in a magnetic field, kinetic inductance stores
energy in the kinetic motion of charge carriers. A unique property
of $L_k$ is its dependence on the current $I$ due to the depairing
effect of moving superconducting electrons, and it plays a crucial
role in the operation of modern superconducting devices. These
include traveling-wave parametric amplifiers
(TWPAs)\cite{Eom_2012,Vissers_2016,Malnou_2021,Faramarzi_2024},
current sensors \cite{Kher_2016}, magnetometers
\cite{Luomahaara_2014}, and even superconducting qubits
\cite{Ku_2010,Faramarzi_2021}. For instance, a kinetic inductance
magnetometer operates by detecting a change in the frequency of an
$LC$ resonator, where the superconductor forms the inductive part,
and Meissner currents induced by an external magnetic field modify
$L$.

In practice, "dirty" superconductors with high normal-state
resistivity $\rho_n$ are commonly used because they provide a
large absolute value of kinetic inductance $L_k \sim \rho_n$ which
exceeds geometric inductance $L_g$. However, a major limitation of
existing materials is the relatively small relative change of
$L_k$ with a current (on the order of a few or tens of percent
\cite{Eom_2012,Vissers_2016,Malnou_2021,Faramarzi_2024,Kher_2016,
Luomahaara_2014,Faramarzi_2021,Andratskii_1976,Healey_2008,Xu_2019,Frasca_2019}).
From a fundamental perspective, the small nonlinearity is
connected with a short superconducting coherence length $\xi$ in
such materials (which is typically a few nanometers) and intrinsic
defects with dimensions larger than $\xi$, which cause premature
instability of the current-carrying state before reaching the
depairing current. As a result, the critical current $I_c$ is far
below the depairing current, the fundamental limit set by the
Cooper pair breaking velocity and where $L_k$ diverges in the
so-called fast relaxation limit
\cite{Andratskii_1976,Anlage_1989,Clem_2012}. Thus, the
experimentally accessible range of supercurrents, where $L_k$
would vary noticeably, remains severely restricted.

To overcome these limitations, recent efforts have focused on
superconducting hybrids: layered structures combining a 'dirty'
superconductor (S) with a low resistive normal metal (N). In such
systems, the proximity effect can strongly modify the
current-momentum relation $I(q)$ \cite{Ustavschikov_2021} and,
consequently, the kinetic inductance $L_k \sim dq/dI$. Due to the
proximity-induced superconductivity in the N layer, the N layer
can sustain a supercurrent that can be even larger than the
supercurrent flowing through the S layer -- it was first
theoretically predicted in \cite{Aslamazov_1986}. This effect is a
consequence of the large ratio of resistivities (diffusion
coefficients) $\rho_S/\rho_N \sim D_N/D_S \gg 1$ of the layers and
the large contribution of the N layer to the total kinetic
inductance. Due to the different diffusion coefficients,
superconductivity in the N layer is suppressed at a lower momentum
$q$ (in the 'dirty' limit, the critical momentum $q_c \sim
1/\sqrt{D}$)\cite{Maki_1963}), while the superconductivity in the
S layer is only weakly affected. This leads to a significant
increase in the total inductance $L_k$ without transition to the
resistive state. Experimentally, considerable increase in the
nonlinearity of $L_k$ upon depositing a thin normal metal layer
was observed in MoN/Cu \cite{Ustavschikov_2021}, NbN/Mo
\cite{Korneeva_2025}, and NbN/Al \cite{Claassen_1999} bilayers.

\begin{figure}[hbtp]%1
\includegraphics[width=0.45\textwidth]{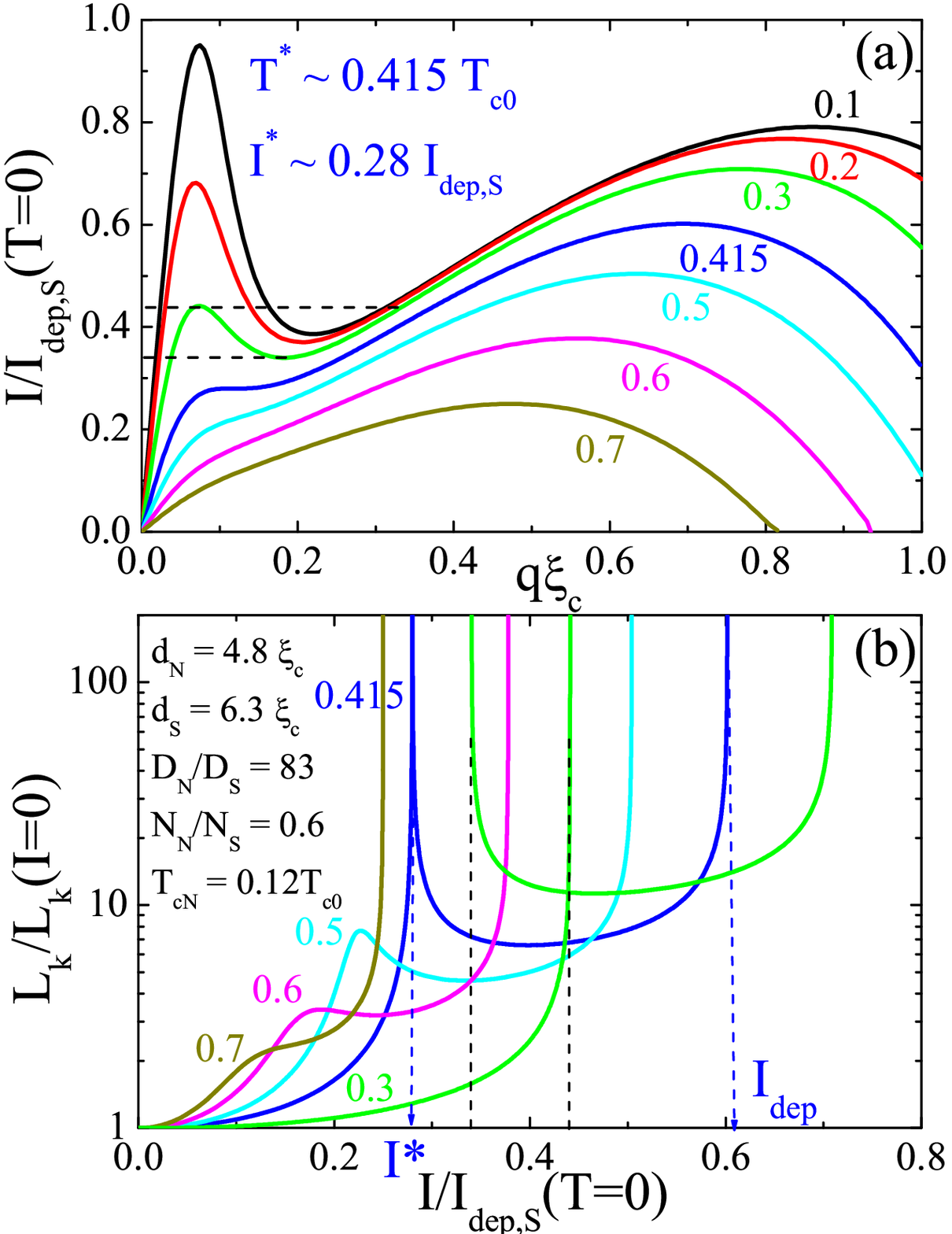}
\caption{(a) Theoretical dependence of the superconducting current
on momentum for an NbN/Al hybrid at different temperatures,
color-coded and measured in units of the critical temperature of
the S layer $T_{c0}$ (here $T_{c0}=10$K and $\xi_c=\sqrt{\hbar
D_S/k_BT_{c0}} \simeq 6.2$ nm for our NbN with $D_S=0.5 cm^2/s$).
(b) Dependence of the kinetic inductance ($L_k \sim dq/dI$) on the
superconducting current at the different temperatures. At $T =
0.415 T_{c0}$ there is a plateau in the dependence $I(q)$, leading
to a peak-like dependence $L_k(I)$. For $T < T^*$, the theory
predicts a two-valued $L_k$ in finite current interval, marked at
$T=0.3 T_{c0}$ by black dashed lines in figures (a,b). In the
inset we show the parameters of the SN bilayer used in the
calculations based on Usadel model ($D_{N,S}$ are diffusion
coefficients, $N_{N,S}$ are densities of states at the Fermi level
in the normal state, $T_{cN}$ is the critical temperature of the
'normal' layer, $I_{dep,S}(T=0)$ is the depairing current of
single S layer.) For the details of the used model see Appendix
B.}
\end{figure}

The effect of the normal layer on the $L_k(I)$ nonlinearity can be
drastically enhanced by choosing the parameters of the S and N
layers such that the $I(q)$ dependence acquires a new qualitative
feature -- the two maxima at low temperatures
\cite{Ustavschikov_2021,Vodolazov_2023} -- see Fig. 1(a). The
maximum at small $q$ corresponds to the suppression of
superconductivity in the N layer, while the maximum at large $q$
corresponds to the suppression of superconductivity in the S
layer. As the temperature increases, the amplitude of the first
maximum decreases due to the reduction of the proximity-induced
superconductivity in the N layer. This leads to the appearance of
a plateau in the $I(q)$ dependence at some temperature $T^*$ and
current $I^*$. This results in a divergence $L_k \sim dq/dI \to
\infty$ as $I \to I^*$ \cite{Vodolazov_2023}, which lies much
below the depairing current of the SN bilayer (the latter
corresponds to the maximal current on the $I(q)$ curve where $L_k$
diverges).

In our work we confirm this theoretical prediction experimentally.
Figure 2(a) shows the current dependence of the total inductance
$L=L_g+L_k$ in the range $0 < I < I_c$ for a NbN/Al with
thicknesses 40 nm and 30 nm respectively and strip length $\ell =
500 \mu m$ and width $w = 1.6 \mu m$, measured at various
temperatures. In the experiment, the impedance of the strip (the
imaginary part of which is equal to $2\pi f L$) was directly
measured at frequencies $f<1$ MHz (the measurement method and the
sample fabrication procedure are presented in the Appendix A). The
largest nonlinearity of $L(I)$ is observed at $T \sim 3$ K, which
is lower than the expected $T^* \sim 4.2$ K from our calculations.
At higher temperatures, the evolution of $L(I)$ qualitatively
follows that of the Usadel model (cf. Figures 1(b) and 2(a)).
However, at lower temperatures, the predicted two-valued behavior
of $L$ in finite current interval (see black dashed lines in Fig.
1(b)) is not observed. Instead, as the temperature decreases, the
amplitude of the peak in $L(I)$ monotonically decreases and
disappears at $T \sim 2.1$ K. At even lower temperatures, the
nonlinearity of $L(I)$ is insignificant.

\begin{figure}[hbtp]%2
\includegraphics[width=0.45\textwidth]{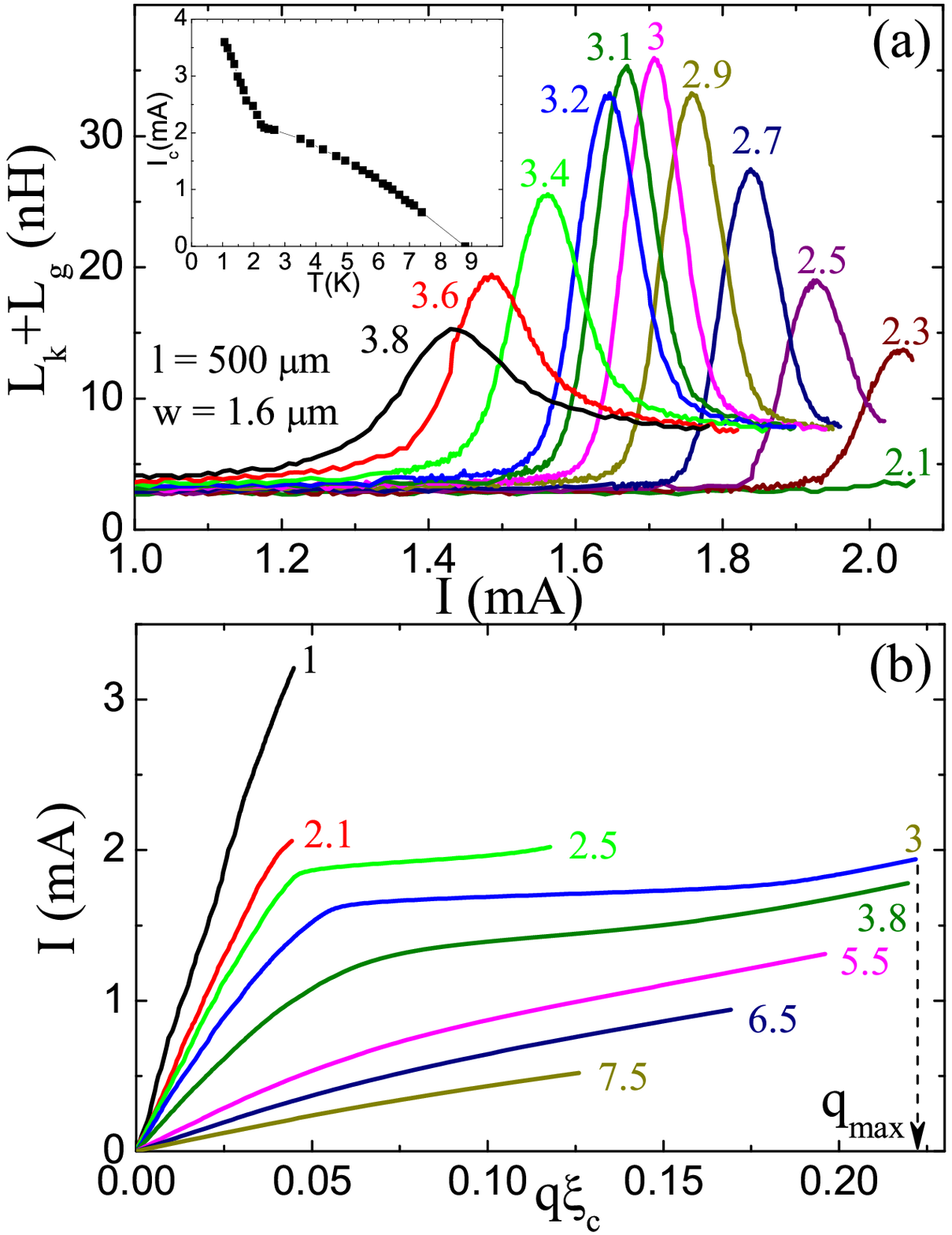}
\caption{((a) Current-dependent inductance of the NbN(40 nm)/Al(30
nm) strip ($w=1.6 \mu m$, $\ell=500 \mu m$) at different
temperatures, indicated by colored numbers. The inset shows the
temperature-dependent critical current of the hybrid strip. (b)
Current-momentum dependence for the same sample.}
\end{figure}

The measured inductance includes a contribution from the geometric
inductance $L_g$. The value $L_g = 2.4 \pm 0.1 \text{ nH}$ for
this sample was determined by comparing (i) the temperature
dependence of $L_k(T)$ obtained from mutual inductance
measurements ($M(T) \sim L_k(T)$) using two coils separated by a
superconducting film (for details, see Appendix A), and (ii) the
temperature-dependent impedance of the superconducting strip (see
Appendix C).

Knowing $L_g$ and $L_k(I)=L(I)-L_g$, the dependence $q(I) =
2e/(\hbar \ell)\int_0^I L_k(I)dI$ and, hence, $I(q)$ can be
calculated (see Fig. 2(b)). Note that the value $q\xi_c = 0.06$ at
which the plateau in the $I(q)$ dependence appears at $T=3$K is
close to the theoretically predicted value $q\xi_c \simeq 0.08$
(compare with Fig. 1(a)). However, the maximal value $q_{max}\xi_c
\simeq 0.22$ is significantly smaller than the theoretical value
$q_{max}\xi_c = q_c\xi_c \simeq 0.7-0.8$ (it corresponds to the
position of the second peak in $I(q)$ -- see Fig. 1(a)). We
attribute this discrepancy to the influence of edge defects, which
leads to vortex entry and motion across the strip -- i.e., a
transition to the resistive state -- at smaller values of $q$ and
current than predicted by theory for a homogeneous sample. The
further decrease of $q_{max}$ with decreasing temperature can be
explained by the appearance of the first peak in the $I(q)$
dependence. Indeed, in the current-biased regime, when the current
exceeds the value corresponding to the first maximum, the hybrid
transitions to a state with a much larger $q$ (see the upper black
dashed line in Fig. 1(a)), and such a $q$ already corresponds to
the resistive state.

The smooth decrease in the width and height of the $L(I)$ peak
with decreasing temperature below 3K may be related to variations
in the strip width along its length. Indeed, our electron
microscopy measurements indicate a non-uniform strip edge with a
characteristic roughness depth of $\sim 100$ nm, which leads to a
variation in the strip width along its length of about $6\%$ and a
similar variation in $I^*$. This results in the appearance of a
plateau in $I(q)$ at different currents for different sections of
the strip. The same reason may also lead to a relatively broad
$L(I)$ peak at $T\sim T^*$. Note that, in combination with the
small experimentally achievable value $q_{max}\xi_c \sim 0.22$,
this may also account for the absence of two-valued $L(I)$ at high
currents. Indeed, upon a slight decrease in temperature below
$T^*$, the two-valued behavior may be smeared out due to width
variations, whereas for a larger deviation from $T^*$, a
transition to a state with $q>q_{max}$ occurs, implying the
presence of a resistive state.

In the inset to Fig. 2(a) we present the temperature dependence of
the critical current of the hybrid strip. It can be seen that
below 2 K, $I_c$ increases sharply. Such a dependence was
previously predicted for superconducting hybrids having two maxima
in the $I(q)$ dependence in Ref. \cite{Ustavshikov_2020}. It is
associated with the appearance of the first maximum in the $I(q)$
dependence, the amplitude of which (i.e., the magnitude of the
critical current) grows rapidly with decreasing temperature, in
contrast to the amplitude of the second peak (or the current at
$q_{max}\xi_c = 0.22$ -- the critical current of the S layer) -- see
Fig. 1(a). That is, such a dependence is a fingerprint of an
$I(q)$ with two maxima.
\begin{figure}[hbtp]%3
\includegraphics[width=0.48\textwidth]{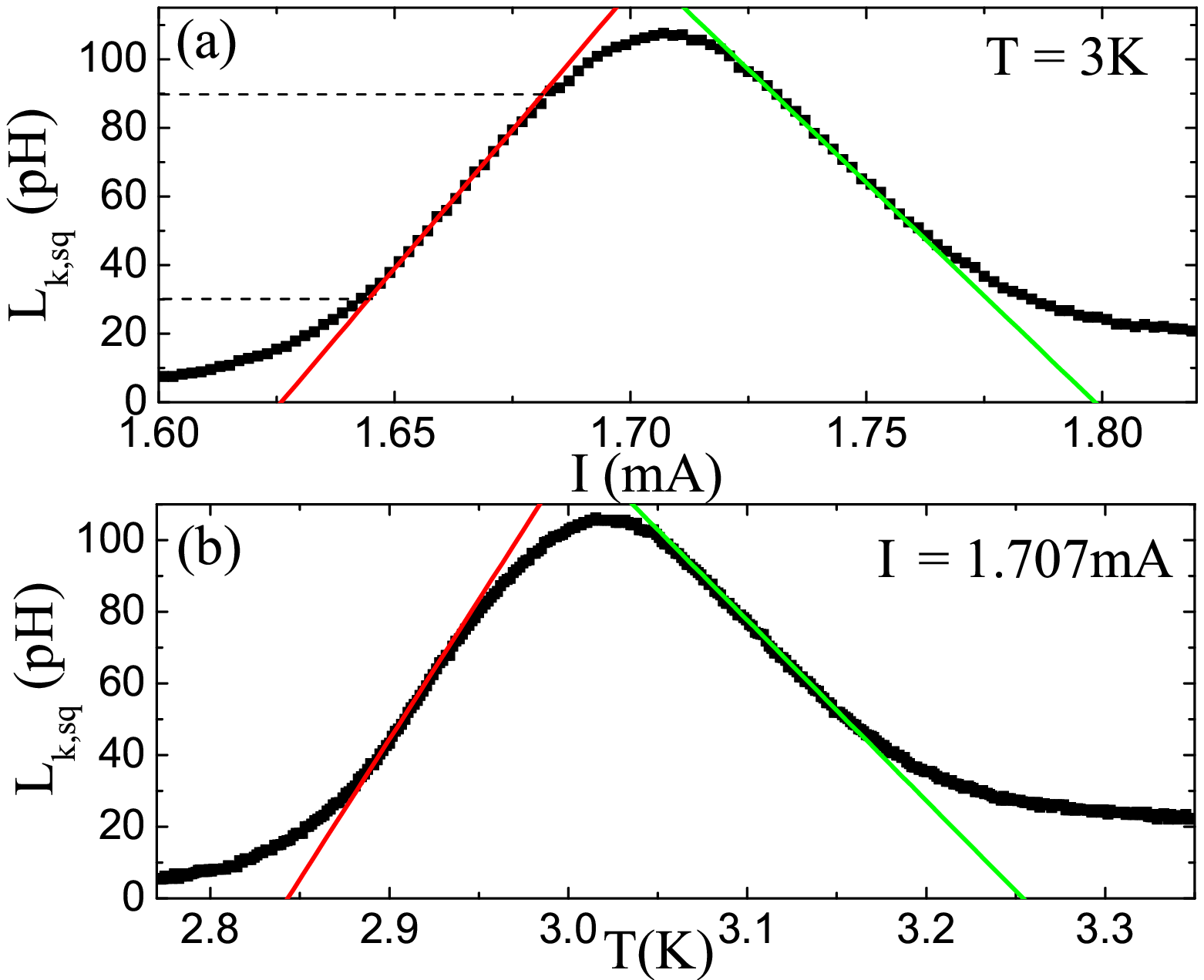}
\caption{(a) Enlarged part of $L_k(I)$ from Fig. 2(a) at $T = 3$
K. Red and green lines indicate nearly linear dependencies with
positive and negative slopes $dL_k/dI$. (b) Dependence of the
kinetic inductance on the temperature of the NbN(40 nm)/Al(30 nm)
strip at $I_{dc} = 1.707$ mA (it corresponds to the maximum of
$L_k(I)$ at $T = 3$ K). Red and green lines indicate nearly linear
dependencies with positive and negative slopes $dL_k/dT$.}
\end{figure}

Figure 3(a) shows, on an enlarged scale, the dependence
$L_{k,\text{sq}}(I)=L_k(I)w/\ell$ (kinetic inductance per square)
at $T = 3$ K. We highlight with a red line a region with a nearly
linear dependence where $L_{k,\text{sq}}$ changes threefold from
30 to 90 pH in a small current range (the total change in
$L_{k,\text{sq}}$ is 50-fold, from $L_{k,\text{sq}}(I=0) = 2 \pm
0.2$ pH to $L_{k,\text{sq}}(I=1.707\text{ mA}) \simeq 107$ pH).
The demonstrated change in kinetic inductance is record-breaking
and can be used, for example, to significantly (by two orders of
magnitude) reduce the length of a travelling wave parametric
amplifier, in present versions of which the change in $L_k$ with
current is only a few percent. This effect can also be used to
create a magnetometer in which a hybrid strip current-biased near
$I^*$ (for example, forming part of a loop presented in
\cite{Vodolazov_2023}) will strongly change its inductance in
response to magnetic field-induced Meissner currents. Another
possible application is the use of an SN strip in TES (transition
edge detector) instead of a SQUID, to measure currents arising
from changes in the resistance of the TES sensitive element.

Note that in addition to the region where $L_{k,\text{sq}}$
increases with increasing current, there is also a region where
$L_{k,\text{sq}}$ decreases with increasing current, and over a
fairly wide range the decrease is close to linear (see the green
line in Fig. 3(a)). To the best of our knowledge, such a type of
$L_{k,\text{sq}}(I)$ dependence has not been discussed in the
literature before.

Figure 3(b) shows the $L_{k,\text{sq}}(T)$ dependence at a fixed
bias current $I = 1.707\text{ mA}$ (at this current,
$L_{k,\text{sq}}$ reaches its maximum value at $T = 3$ K). This
result indicates that a current-biased hybrid strip can serve as a
calorimeter/thermometer or electromagnetic radiation detector. For
example, for the green line, a temperature increase of 0.1 K leads
to a twofold decrease in inductance. In these measurements for the
strip with aspect ratio $\ell/w \simeq 312$, we use a frequency $f
= 0.3$ MHz and an amplitude of ac current $I_{ac} = 30 \mu A$.
With an intrinsic noise of our lock-in amplifier of $\sim 20$ nV,
we have an uncertainty in the determination of $L_{k,\text{sq}}$
of about 1 pH. With $dL_{k,\text{sq}}/dT = -0.5$ pH/mK,
corresponding to the green line in Fig. 3(b), we would measure a
change in the temperature of our NbN/Al strip of 2 mK. Increasing
the ratio $\ell/w$ and/or the frequency allows one to increase the
sensitivity of such a kinetic inductance thermometer. We should
note that estimates made in \cite{Vodolazov_2023} show that with
an optimal choice of superconductor parameters (in particular, a
low $T_{c0} \lesssim 0.5$ K is required to ensure $T^* \sim 20$
mK), a single-photon detector in the GHz range can be realized.

In Ref. \cite{Vodolazov_2023}, it was theoretically found that at
currents $I \gtrsim I^*$ in an SN strip, the gapless regime is
realized. Although the gap is formally equal to zero across the
whole thickness, in the S layer the density of states is strongly
suppressed with respect to the normal state value up to an energy
$\sim k_BT_{c0}$ (see Fig. 2 in \cite{Vodolazov_2023}). This makes
it plausible that the quality factor of resonators made of a
current-biased SN strip (or bridge) will be reasonably large.
Furthermore, only a part of the resonator can be made from a
highly sensitive SN hybrid, while the rest could be made from an S
material, as proposed in Ref. \cite{Vodolazov_2023} (see Fig. 1
therein).

We stress that the maximal nonlinearity of $L_k$ is reached at a
current $I = I^*$ that is well below the critical current. This is
in strict contrast with an ordinary superconductor, where the
nonlinearity of $L_k$ increases when approaching $I_c$, and $L_k$
would diverge if one could reach the depairing current. We expect
that by improving the quality of the strip edge (using electronic
lithography instead of photolithography), one may further increase
$I_c$ and move it further from $I^*$.

Similar results were found for NbN(40 nm)/Al(30 nm) strips with
different widths (see Appendix D). Variation in width leads to a
small variation (within 0.1 K) in the temperature $T^*$, while the
current $I^*$ scales almost linearly with $w$, as expected from
theory (the current in Fig. 1 is scaled in units of the depairing
current of the S layer, and $I_{dep,S}\sim w$ when $w$ is less
than the Pearl penetration length).

As the theory predicts \cite{Vodolazov_2023}, the values of
current $I^*$ and temperature $T^*$ depend on the material
parameters of the SN hybrid. The only necessary condition for
their existence is the ratio $\rho_S/\rho_N \gg 1$. To have low
$T^*$, one may use a dirty superconductor with low $T_{c0}$, like
granular Al ($T_{c0} \sim 2$ K) \cite{Levy-Bertrand_2019}, TiN ($0
< T_{c0} < 5$ K) \cite{Leduc_2010}, PtSi ($T_{c0} \sim 1$ K)
\cite{Szypryt_2016}, or Hf ($T_{c0} \sim 0.4$ K)
\cite{Coiffard_2020}, and low-resistive metals such as Au or Cu.
Large $T^*$ could be provided by superconducting materials like
NbN with $T_{c0} \sim 15$ K \cite{Kamlapure_2010}, NbTiN ($T_{c0}
\simeq 13$ K \cite{Pratap_2023}), or MgB$2$ ($T_{c0} \sim 18$ K
\cite{Shibata_2010}).

To conclude, we report a giant (50-fold) current-induced variation
of the kinetic inductance $L_k$ in superconducting NbN(40
nm)/Al(30 nm) hybrid strip. The major change in $L_k$ occurs in a
narrow current interval near $I^*$ (lying below $I_c$) and in a
temperature interval of about $0.5$ K centered at $T^* \approx 3$
K. We found that $I^*$ scales linearly with the strip width, in
agreement with theoretical predictions. We attribute the absence
of predicted two-valued behavior of $L$ at $T < T^*$ and the
relatively wide peak in the $L(I)$ dependence to the low quality
of the strip edges, which leads to width variations and vortex
penetration at small Cooper pair momentum. Our result shows that
this and similar superconducting hybrids are promising candidates
for new types of highly sensitive superconducting detectors of
electromagnetic radiation, calorimeters, thermometers, parametric
amplifiers, and magnetometers.

\begin{acknowledgments}

The study was supported by project No. 125020501540-9 of the
Ministry of Education and Science of the Russian Federation.
Fabrication and technology characterization were carried out at a
large-scale facility complex for heterogeneous integration
technologies and silicon-carbon nanotechnologies.
\end{acknowledgments}

\appendix

\section{Experimental details}

The samples were fabricated using thin-film planar technology.
Low-resistivity silicon wafers, pre-oxidized at $1000^{\circ}$C in
a wet oxygen atmosphere, were used as substrates. The NbN-Al
bilayer system was deposited in situ using DC magnetron
sputtering. First, the NbN film was deposited in a mixture of Ar
and N$2$ gases onto substrates preheated to $350^{\circ}$C,
followed by the deposition of the Al film in an Ar atmosphere at
room temperature without breaking the vacuum. The thickness of the
superconducting layer is $d_{NbN}=40$ nm, while the thickness of
the aluminium layer is $d_{Al}=30$ nm, as confirmed by X-ray
reflectometry (XRR). For transport measurements, strips were made
by photolithography and Ar+ ion etching using a Plasmalab 80 plus
(Oxford Instruments) equipped with inductive plasma sources.

Resistance vs temperature measurements for the unpatterned films
and strips were performed in a Gifford-McMahon cryocooler  with a
minimum achievable temperature of 2.6 K. We used 2-probe
measurement with Keithley 2460 SourceMeter to bias the sample with
1-10$~\mu$ DC current and read voltage. Measurements in wide
temperature range (15 - 300 K) were performed during cooling and
warming cycles of the cryocooler, while precise measurements in
2.6 - 15 K range were performed with the built-in heater on the
cryocooler's cold plate and operated with the LakeShore 325
temperature controller.

\begin{table}
\centering \caption{Parameters of the NbN(40nm)/Al(30nm) strips.}
\label{Table_1}
\begin{tabular}{ccccccc}
\hline
Sample & $\ell$,$\mu$m & $w$,$\mu$m & $L_g$, nH & $L_{k,sq}(1K)$, pH & $\Lambda(1K)$, $\mu$m & $T_c$,K   \\
\hline
S$_{\text{med}}$ & 500 & 1.6 & 2.4 $\pm$ 0.2 & 1.1 $\pm$ 0.1 & 0.9 & 9.5   \\
S$_{\text{wide}}$ & 3000 & 3.5 & 2.2 $\pm$ 0.2 & 1.1 $\pm$ 0.1  & 0.9 & 9.5  \\
S$_{\text{narr}}$ & 50 & 0.6 &  &  &  &  9.4 \\
\hline
\end{tabular}
\end{table}

The current-voltage characteristics of NbN/Al strips were measured
using the four-probe method. From these measurements, we extracted
the temperature-dependent critical current. The found $I_c(T)$
dependencies vary slightly from sample to sample. Therefore we
choose typical samples with different geometrical parameters and
perform inductive measurements using these stripes. The parameters
of studied NbN(40 nm)/Al(30 nm) strips are listed in Table 1. We
did not measure $L(T)$ for sample S$_{\text{narr}}$ due to its
complicated geometry; therefore, we could not determine $L_g$ and
$L_{k,sq}(1\ \mathrm{K})$. The critical temperature $T_c$ is
defined as the temperature at which the resistance equals half of
the resistance at 20 K. A single NbN layer has a critical
temperature $T_{NbN} = 10$ K and a resistivity $\rho_{NbN}=230\
\mu \Omega \cdot \mathrm{cm}$ at $T=20$ K. The resistivity of
aluminium is $\rho_{Al}=4.6\ \mu \Omega \cdot \mathrm{cm}$ at
$T=20$ K. This gives us the desired large ratio
$\rho_{NbN}/\rho_{Al} = 50$.

The impedance $Z = Z_{re} + iZ_{im}$ of the NbN/Al strips was
measured using a DXA-003 DSP lock-in amplifier. An excitation
signal with a frequency $f = 0.3 - 1$ MHz and a voltage of 20 --
100 mV from the DXA-300's internal generator was supplied through
a 2-k$\Omega$ resistor. The signal from the sample was applied to
the differential input of the DXA-300 through the central
electrodes of two coaxial cables. These measurements allowed us to
determine the temperature and current dependence of the inductance
$L = L_k + L_g = Z_{im}/(2\pi f)$. We verified that, in the
frequency and voltage range used, $Z_{im}$ is proportional to $f$
and that it does not depend on the amplitude of the excitation
signal.

In addition, we found the $L_k(T)$ dependence using the two-coil
technique \cite{Claassen_1997}. In this method, one measures the
mutual inductance $M$ between two coils when the tested
superconducting sample is placed between them. As the temperature
is lowered, the screening properties of the superconducting film
increase and $M$ drops. In our case, the diameter of the coils (2
mm) and their height (4 mm), as well as the separation between the
coils (1 mm), are much smaller than the lateral size of the sample
(1 cm). In addition, the total thickness of the NbN/Al bilayer is
much smaller than the London penetration depth for both NbN and
Al. Under these conditions, the measured mutual inductance $M$ is
proportional to the Pearl penetration length $\Lambda$
\cite{Claassen_1991,Vodolazov_2018} and, hence, to $L_k$: $M \sim
\Lambda \sim L_k$, except in the vicinity of $T_c$ where $L_k$
diverges.

Impedance, two-coil and part of transport measurements are
performed in a cryostat using a Cryomesh Pulse Tube 410 with a
minimum achievable temperature of 1.0 K. In impedance measurements
we added a shunt with a resistance of 1$\Omega$ to prevent our
NbN/Al strip from switching to the normal state due to extrinsic
current noise when approaching the critical current. We verified
that its presence does not affect the $L(I)$ dependence (we
performed test measurements with and without the shunt).

\section{Model}

To calculate the dependence $I(q)$ for an SN bilayer, we use the
one-dimensional Usadel equation for the normal $g$ and anomalous
$f$ quasiclassical Green functions. With the standard angle
parametrization $g = \cos\Theta$ and $f = \sin\Theta ,
e^{i\varphi}$, the Usadel equation in different layers can be
written as
\begin{eqnarray}
\hbar D_{S,N}\frac{\partial^2\Theta_{S,N}}{\partial
z^2}-\left(\hbar\omega_n+D_{S,N}\hbar q^2\cos
\Theta_{S,N}\right)\sin\Theta_{S,N} \nonumber
\\
+2\Delta_{S,N}\cos\Theta_{S,N}=0,
\end{eqnarray}

where the subscripts S and N refer to the superconducting and
normal (or having a lower critical temperature) layers,
respectively. Here $D_{S,N}$ is the diffusion coefficient for the
corresponding layer, $\hbar \omega_n = \pi k_B T (2n+1)$ are the
Matsubara frequencies (where $n$ is an integer), $\hbar q = \hbar
\nabla \varphi$ is the momentum of the Cooper pairs, and $\varphi$
is the phase of the order parameter (it is the same in both
layers). $\Delta_{S,N}$ is the superconducting order parameter,
which satisfies the self-consistency equation
\begin{equation}
\label{self-cons} \Delta_{S,N}
\ln\left(\frac{T}{T_{c0,cN}}\right)=2\pi k_B T\sum_{\omega_n
>0} \left(\sin\Theta_{S,N} - \frac{\Delta_{S,N}}{\hbar\omega_n}\right),
\end{equation}
where $T_{c0}$, $T_{cN}$ are the critical temperatures of single S
and N layers, respectively. These equations are supplemented by
the Kupriyanov-Lukichev boundary conditions on the SN interface
\cite{JETP-1988}
  \begin{equation}
    D_S\frac{d\Theta_S}{dz}=D_N\frac{d\Theta_N}{dz}
   \end{equation}
For simplicity, we consider the case with zero barrier between the
layers and continuous $\Theta_S = \Theta_N$ at the SN interface.
For interfaces with vacuum, we use the boundary condition
$d\Theta_{S,N}/dz = 0$.

To calculate the current density $j$ and current $I=w\int jdz$ we
use the standard expression
\begin{eqnarray}
j(z)=\frac{2\pi k_BT}{e\rho_{S,N}}q\sum_{\omega_n >
0}\sin^2\Theta_{S,N}
\end{eqnarray}
where $\rho_{S,N}=2|e|D_{S,N}N_{S,N}$ are the residual
resistivities and $N_{S,N}$ are density of states of electrons per
one spin at the Fermi level in the normal state, in the
corresponding layer.

Equations (B1) -- (B3) are solved numerically using an iteration
procedure. For an initial distribution $\Delta(z) = \text{const}$
and a chosen $q$, we solve Eqs. (B1), (B2) (in the numerical
procedure, we used Newton's method combined with the tridiagonal
matrix algorithm). The found solution $\Theta(z)$ is inserted into
Eq. (B3) to find $\Delta(z)$, and then the iterations are repeated
until the relative change in $\Delta(z)$ between two iterations
does not exceed $10^{-8}$. Length is normalized in units of $\xi_c
= \sqrt{\hbar D_S / k_B T_{c0}}$, energy is normalized in units of
$k_B T_{c0}$, and current is normalized in units of the depairing
current $I_{\text{dep},S}$ of a single S layer with thickness
$d_S$ at zero temperature. The grid step in the S and N layers is
$\delta z = 0.1 \xi_c$. In the calculations, we used the following
parameters: $d_S = 6.3 \xi_c$, $d_N = 4.8 \xi_c$, $D_N/D_S = 83$,
$N_N/N_S = 0.6$. These values correspond to the parameters of our
NbN(40 nm)/Al(30 nm) bilayer with a typical value $D_S =
D_{\text{NbN}} = 0.5\ \text{cm}^2/\text{s}$ and $\xi_c = 6.2$ nm
(this can be seen from a comparison of the calculated $L_k(T)$
with the experimental results -- see the inset of Fig. B1).

With known $I(q)$, we find the kinetic inductance of the strip
\begin{equation}
L_k=\frac{\hbar \ell}{2e}\frac{dq}{dI}.
\end{equation}

Eq. (B5) follows from the voltage drop $V$ along the
superconducting strip having kinetic inductance $L_k$
\begin{equation}
V=L_k \frac{dI}{dt}
\end{equation}

and Josephson relation
\begin{equation}
V=\frac{\hbar}{2e} \frac{d \delta \phi}{dt}=\frac{\hbar \ell}{2e}
\frac{dq}{dt}.
\end{equation}

where $\delta \phi = q \ell$ is a phase difference along the
strip.

In an SN strip, as in an ordinary S strip, at small currents (and,
hence, small momentum of superconducting electrons $q$), one may
expand $I(q)$ in a Taylor series.
\begin{equation}
I(q) \sim Aq-Bq^3,
\end{equation}
where $A$ and $B$ are temperature- and sample-dependent
coefficients, and the term proportional to $q^2$ is absent because
$I(-q)=-I(q)$. The cubic term provides a quadratic dependence
\begin{equation}
L_k(I)=L_k(0) + C I^2.
\end{equation}
Indeed,
\begin{equation} L_k(I) \sim
\frac{dq}{dI} \sim \frac{1}{A-3Bq^2}\sim
\frac{1}{A}+\frac{3Bq^2}{A^2} \sim L_k(0)+CI^2,
\end{equation}

with $L_k(0) \sim 1/A$ and $C=3B/A^4$. However, at large $q$, Eq.
(B9) no longer holds, especially near the plateau or maximum of
$I(q)$, where $L_k$ diverges.

In ordinary superconductors, the critical current $I_c$ rarely
reaches even half of the depairing current, which means that $q$
is far from the value $q_{dep}$, near which the dependence $I(q)$
is most nonlinear and reaches its maximal value, known as the
depairing current $I_{dep}=I(q_{dep})$. This leads to a relatively
small current-induced variation of $L_k$: $L_k(I_c)-L_k(0) \ll
L_k(0)$ and $L_k(I)$ is described by Eq. (B9).

In the case of an SN bilayer with proper parameters, it is
possible to have a large nonlinearity of $I(q)$ at relatively
small $q$ and a highly nonlinear $L_k(I)$ not only near depairing
current, but also near the current $I^*\ll I_{dep}$ (see Fig. 1).

\begin{figure}[hbtp]%4
\includegraphics[width=0.48\textwidth]{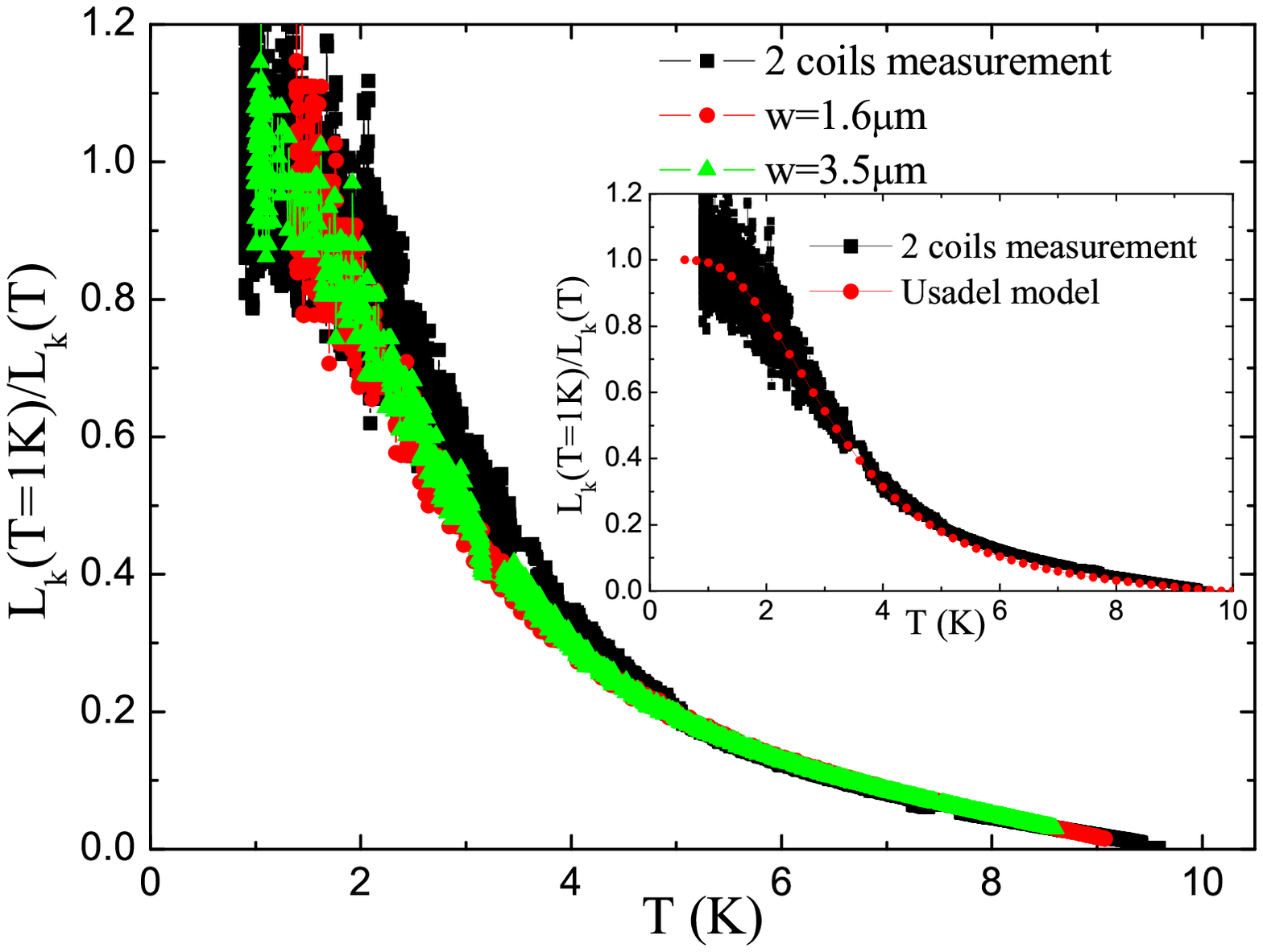}
\caption{Dependence of the kinetic inductance of
NbN(40nm)/Al(30nm) film on temperature, measured using two-coil
technique. On the same figure we show $L_k(T)$ for samples
S$_{\text{med}}$ and S$_{\text{wide}}$ which has been extracted
from impedance measurements with $L_g$ as the only fitting
parameter. In the inset we compare theoretical and experimental
$L_k(T)$ dependencies.}
\end{figure}

\section{Temperature dependence of $L_k$}

Figure 4 shows the dependencies of $L_k(T)$ obtained using the
two-coil technique and temperature-dependent impedance. For the
strips, the geometric inductance $L_g$ was used as the only
fitting parameter, the value of which is presented in Table 1. The
resulting $L_{k,\text{sq}}(1\text{ K})$ values for the different
strips are given in Table 1.

\section{$L(I)$ dependencies for wider and narrower strips}

Figures 5 and 6 show, respectively, the dependence of $L(I)$ at
various temperatures and the $I_c(T)$ dependence for NbN(40
nm)/Al(30 nm) strips with different widths. Qualitative agreement
with the results obtained for sample S$_{\text{med}}$ is observed,
with a small variation (within 0.1 K) in the temperature $T^*$. At
the same time, the current $I^*$ scales almost linearly with the
strip width, as expected from theory.

For sample S$_{\text{wide}}$, the maximum change in $L_k$ is about
of 40 at $T = 3.1$ K. Note that for this sample, the strip width
slightly exceeds the Pearl penetration depth
$\Lambda(3.1\text{K})\simeq 2.1 \mu \text{m}$, which leads to a
somewhat non-uniform current distribution across the strip at low
currents.

For sample S$_{\text{narr}}$, the change in $L$ is much smaller.
This sample has a more complex geometry, where the total
inductance includes a significant contribution from relatively
long and wide banks connected to the strip (their contribution to
$L$ is approximately $54 L_{k,\text{sq}}(I=0)$). Moreover, for
this sample, the width variation along the strip is about $16\%$,
which further broadens the peak in $L(I)$ and reduces its
amplitude.
\begin{figure}[hbtp]%5
\includegraphics[width=0.48\textwidth]{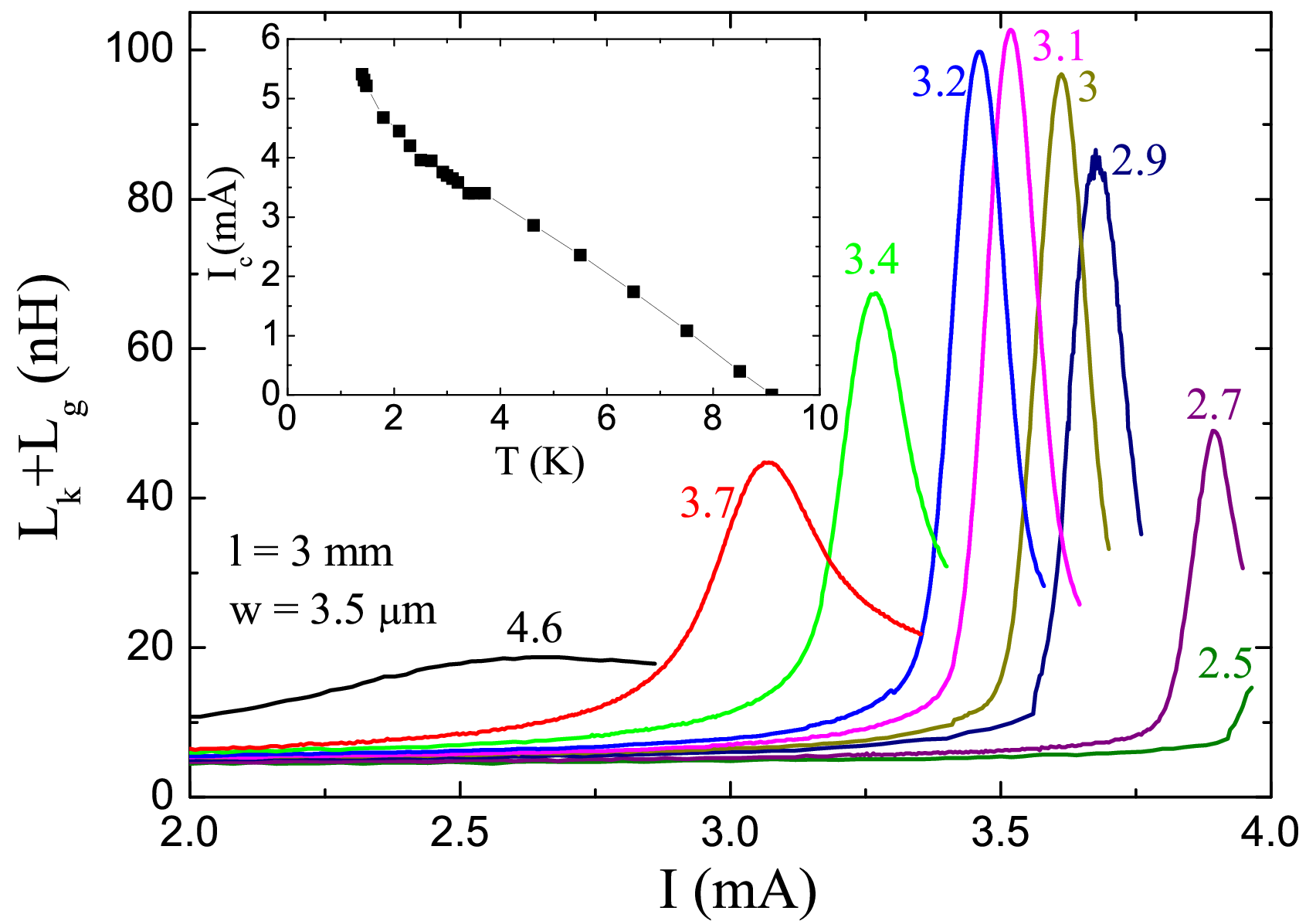}
\caption{Current-dependent inductance of a wide NbN(40 nm)/Al(30
nm) strip ($w = 3.5\ \mu\text{m}$, $\ell = 3 \text{mm}$) at
different temperatures (color-coded). The inset shows the
temperature-dependent critical current of the sample.}
\end{figure}

\begin{figure}[hbtp]%6
\includegraphics[width=0.48\textwidth]{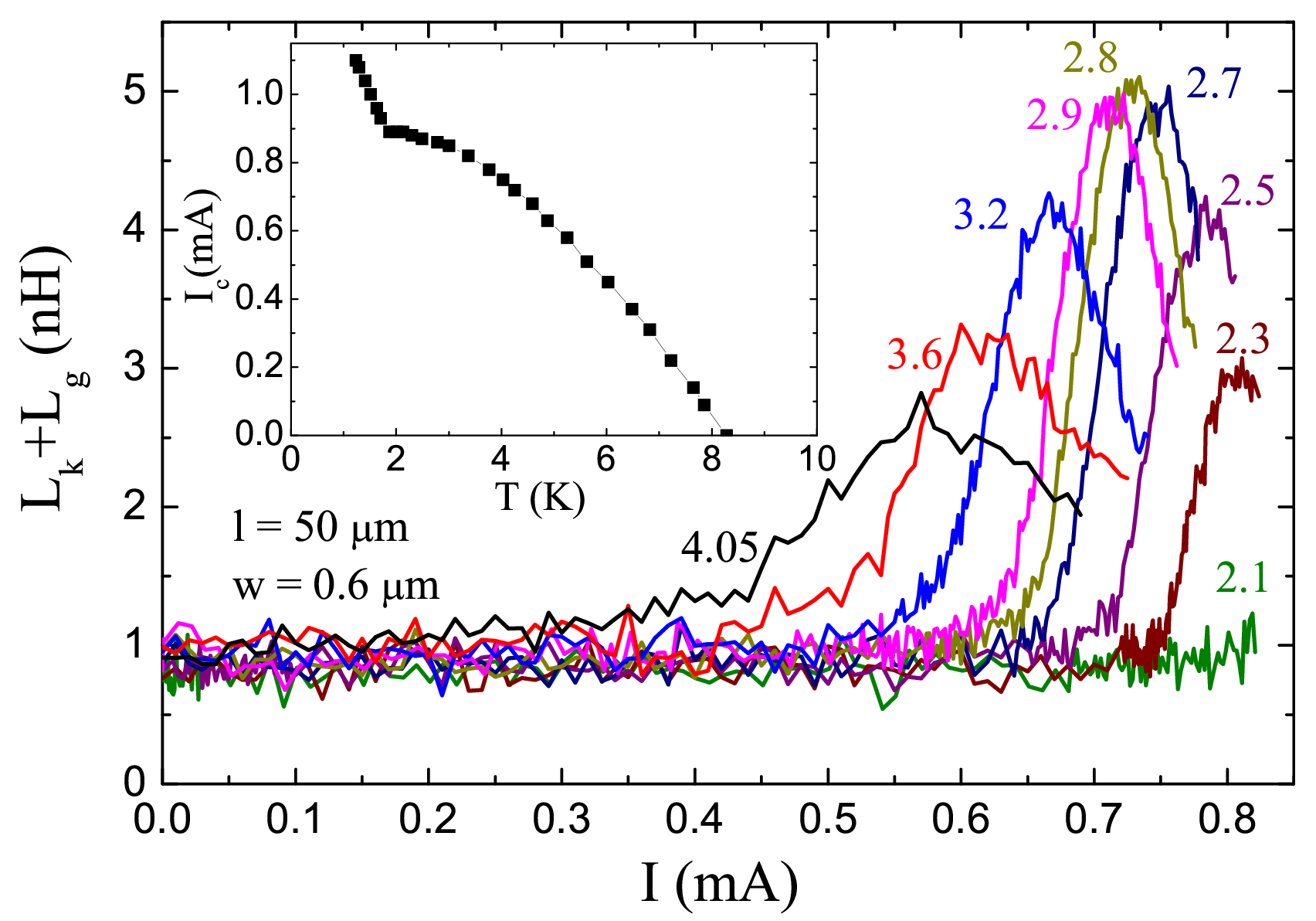}
\caption{Current-dependent inductance of a narrow NbN(40 nm)/Al(30
nm) strip ($w = 0.6 \mu\text{m}$, $\ell = 50 \mu \text{m}$) at
different temperatures (color-coded). Dependence $L(I)$ is noisier
compared to other samples due to the smaller ratio $\ell/w$,
leading to a smaller signal-to-noise ratio. The inset shows the
temperature-dependent critical current of the sample.}
\end{figure}

\end{document}